**Title**: Leveraging contact network structure in the design of cluster randomized trials

**Running Head:** Network structured designs for cluster trials

**Authors**: Guy Harling,[1] Rui Wang,[2,3] Jukka-Pekka Onnela,[2] Victor De Gruttola.[2]

1. Department of Global Health and Population, Harvard T.H. Chan School of Public Health

2. Department of Biostatistics, Harvard T.H. Chan School of Public Health

3. Department of Medicine, Brigham and Womens Hospital

**Address for correspondence**: Guy Harling, Department of Global Health and Population, Harvard T.H. Chan School of Public Health, 1639 Tremont Street, Boston, MA 02120. +1 617 432 1232

gharling@hsph.harvard.edu.

**Financial support**: This research was supported by the National Institutes of Health [grant numbers R37 AI51164, R01 AI24643].

**Word counts**:

Abstract:     372

Article:      4000

# ABSTRACT

**Background**: In settings like the Ebola epidemic, where proof-of-principle trials have succeeded but questions remain about the effectiveness of different possible modes of implementation, it may be useful to develop trials that not only generate information about intervention effects but also themselves provide public health benefit. Cluster randomized trials are of particular value for infectious disease prevention research by virtue of their ability to capture both direct and indirect effects of intervention; the latter of which depends heavily on the nature of contact networks within and across clusters. By leveraging information about these networks – in particular the degree of connection across randomized units – we propose a novel class of connectivity-informed cluster trial designs that aim both to improve public health impact (speed of control l epidemics) while preserving the ability to detect intervention effects.

**Methods**: We consider cluster randomized trials with staggered enrollment, in each of which the order of enrollment is based on the total number of ties (contacts) from individuals within a cluster to individuals in other clusters. These designs can accommodate connectivity based either on the total number of inter-cluster connections at baseline or on connections only to untreated clusters, and include options analogous both to traditional Parallel and Stepped Wedge designs. We further allow for control clusters to be "held-back" from re-randomization for some period. We investigate the performance of these designs in terms of epidemic control (time to end of epidemic and cumulative incidence) and power to detect vaccine effect by simulating vaccination trials during an SEIR-type epidemic outbreak using a network-structured agent-based model.

**Results**: In our simulations, connectivity-informed designs lead to lower peak infectiousness than comparable traditional study designs and a 20% reduction in cumulative incidence, but have little impact on epidemic length. Power to detect differences in incidence across clusters is reduced in all connectivity-informed designs. However the inclusion of even a brief "holdback" restores most of the power lost in comparison to a traditional Stepped Wedge approach.

**Conclusions**: Incorporating information about cluster connectivity in design of cluster randomized trials can increase their public health impact, especially in acute outbreak settings. Using this information helps control outbreaks – by minimizing the number of cross-cluster infections – with modest cost in power to detect an effective intervention.

**BACKGROUND**

Vaccine study designs typically focus on ensuring sufficient power to detect effects for the product under consideration [1]. In an epidemic setting, however, rapid disease control may also be of vital import. This imperative can be seen in use of a "Ring" vaccination trial – a method previously employed to control smallpox and foot-and-mouth [2, 3] – against Ebola in Guinea [4, 5]. In the Ebola trial, vaccination of all contacts was immediate in the intervention arm, and delayed for three weeks in the control arm [6]. While Ring Vaccination is likely to increase the speed of epidemic control, it requires considerable resources to conduct detailed contact tracing, and to maintain an active presence in all clusters.

Conducting statistical inference in the context of vaccine (and other infectious disease) trials is complicated by "dependent happenings" – where one's risk of infection depends on the health status of others – which may lead to interference between treatment and control groups [7, 8]. Cluster randomized trials (CRTs) allow the estimation of combined direct (benefit of your vaccination) and indirect (benefit of others' vaccination) effects. In the commonly-used "Parallel" design, clusters are pair-randomized to treatment or control, and then followed-up for a pre-determined period of time; in settings where use of a control arm is considered unethical, an alternative "Stepped Wedge" design treats all clusters sequentially in a randomized order. In this latter design effect can be measured through some combination of between- and within-cluster comparisons, accounting for the presence of temporal effects [9].

Standard Parallel and Stepped Wedge designs benefit from cluster randomization to prevent possible confounding by underlying differences among clusters [10], and in the case of Stepped Wedge designs also over time [11]. Concerns regarding risk factor imbalance in CRTs have historically led to matching designs, in which pairs or groups of clusters are chosen at baseline and then randomized to treatment or control, as potentially providing some increase in study efficiency [12]. Such designs are particularly attractive in the context of infectious diseases, given the likelihood of considerable heterogeneity in outcomes across clusters [13].

CRT study designs generally minimize contamination between study arms that arises when individuals in different arms have contact with one another [10], but this is not also feasible. For example, in CRTs for HIV prevention, individuals in one cluster may have partners in another [14]; in Ebola vaccine trials, infected individuals may travel for care to homes or hospitals within another trial cluster [15]. In an epidemic setting, the degree of connection between clusters is likely to predict outcomes of interest, including outbreak timing within a cluster and epidemic size. Taking between-cluster connectivity into account can therefore aid in matching.

The purpose of a vaccine is to convert potentially-infectious network ties (i.e. the direct connection between infectious person $i$ and susceptible person $s$) into non-functional ones [16-18]. This conversion can be achieved by successfully vaccinating either end of the tie. Hence, vaccination acts by removing ties from a graph that represent potentially infecting pathways within a population. Contamination can be also conceptualized as a network problem – since ties between individuals across clusters can lead to the spread of either infections or behaviors from one study arm to another, thereby attenuating the impact of randomization.

We propose a novel class of CRT study designs that make use of information about the network connectivity between study clusters; we show that these designs can reduce the number of new infections more rapidly than standard designs while still allowing for the evaluation of intervention effectiveness. Our approach requires less intensive follow-up than does a Ring Vaccination approach. We focus exclusively on trials with staggered implementation in order to mimic urgent settings in which epidemic control is a priority. We investigate the performance of these designs by simulating vaccination trials during an Ebola-like epidemic and evaluate both epidemic parameter values and power to detect an effect of the vaccine under various designs.

**METHODS**

**A class of connectivity-informed cluster trial designs**

Connectivity defines how individuals or groups in a network are linked to one-another. For sexually transmitted infections, ties are sexual acts; for hemorrhagic fevers, physical contacts; for behavior change interventions, conversations. In our study, we consider ties measured prior to study commencement; for each cluster we calculate the absolute number of ties from members of the cluster to members of all other clusters. We then rank clusters from most- to least-connected in one of two ways: the Static Rank approach, where the ranking is conducted only once at baseline; and the Adaptive Rank approach, where untreated clusters are re-ranked at each vaccination time point based only on their connectedness to remaining untreated clusters. Both approaches are based on the idea that a cluster's connectivity to all other clusters is related to its tendency to transmit infections; hence treating more connected clusters earlier may slow epidemic spread. We outline the proposed study designs in Table 1.

Within the Static Rank approach, we consider several different designs: First, a "Strict Order" design which rolls-out treatment from most- to least-connected cluster; this non-randomized approach provides an upper bound on how fast the epidemic might be controlled using between-cluster connectivity information. Second, a "Fuzzy Order" design which randomizes the two most-connected clusters to treatment and control status at the time of study origin (step 1). At the next time of randomization, the control cluster from step 1 and the next most-connected cluster are randomized. This process is repeated until the final untreated cluster remains, which is then assigned to treatment. The Fuzzy Order design can be generalized to a "Fuzzy Order Holdback-$h$" design, in which the control cluster at each time point is held-back from randomization for $h$ vaccination time points: if h=1, the control cluster from step 1 would be re-eligible for

Page 6

randomization at step 3; for h=2, at step 4, etc. The only difference between Static Rank designs is the order in which clusters are treated. We illustrate how these Static Rank designs operate in Figure 1.

We also propose a "Ranked Parallel" design, which randomizes the two most-connected clusters to treatment and control condition at study origin. At step 2, the randomization process is repeated using the third- and fourth-most connected clusters, and so on until half the clusters are assigned to treatment. At this point the collected data are analyzed, and if the treatment proves protective, it is rolled out to the untreated clusters, starting with the most connected.

The Strict, Fuzzy and Holdback Ordered designs can all also use an Adaptive Rank approach, although the Ranked Parallel design cannot – since all randomization pairs must be specified simultaneously. The Adaptive Rank approach minimizes the treatment of people already connected to treated subjects, by including only unvaccinated clusters in between-cluster connectivity: since successful vaccination removes ties between vaccinated and unvaccinated subjects, there is no further benefit to vaccinating the unvaccinated member of a connection. At the cluster level, this implies that vaccinating a cluster that is highly-connected to already-treated clusters is likely to provide less population-level benefit than vaccinating one largely connected to untreated clusters.

Removing treated clusters from the set of potentially transmitting ties can lead to significant re-rankings (see Figure 1B), and thus more rapid epidemic control. The cost of the Adaptive Rank approach is its requirement for more detailed data than the Strict Rank approach: the latter requires only an ordering of clusters by their overall connectivity ($K$ quantities in a study of $K$ clusters), whereas the former requires a measure of connectivity for every pair of clusters ($\binom{K}{2} = \frac{1}{2}K(K-1)$ quantities).

**Simulation studies**

We generate a community-structured population using a standard stochastic block model with $K$ blocks or clusters, each block consisting of $N$ nodes (individuals) [19]. Mean degree for each individual is set at 5.5. In the baseline simulation, we assume that within each cluster, ties are distributed randomly. Half of all clusters are designated to have higher external connectivity: in these clusters individuals have a number of between-cluster ties drawn from a normal distribution with a mean of 1 versus 0.5; all clusters have a standard deviation of 0.5.

We simulate an epidemic on the network graph of the structured population, based on Ebola – as a recent example of a disease requiring urgent vaccine trials, and a member of the viral hemorrhagic fever group quite likely to cause future severe, acute outbreaks [20]. We use a state transition model with six states: Susceptible, Exposed, Infectious, Hospitalized, Funeral, and Removed (see **Supplementary Figure 1**) [21]. Parameter values for the simulation are calibrated such that progression times between states and the basic reproductive number ($R_0$, the average number of new infections caused by an infectious individual in a fully susceptible population) are roughly equal to those observed for Ebola. These values were not optimized to historical data since the simulation is intended for design comparison, not Ebola epidemic prediction.

To compare study designs, we first generate a network realization from the stochastic block model. We then simulate nine epidemics on the network using the above six-state epidemic model; one for each trial design. We initialize the epidemic model by randomly selecting four nodes at the beginning of the simulation to be infected and use the same initial condition for each study design (and for the reference simulation involving no intervention at all). In each case, the epidemic is propagated on the underlying network using daily time steps, and allowed to run for 42 days (six weeks) from initial introduction of infection. If all nine epidemics have substantial ongoing transmission at this point – specifically the effective reproductive number ($R_e$, the average number of new infections actually caused by each infectious person) is greater than one in week six – then we begin the trial; otherwise we discard this network realization.

We simulate vaccinating one cluster every seven days; even in Parallel trials, roll-out is likely to be staggered for logistical reasons – as for example in both Sierra Leone's and Liberia's Ebola vaccine trials [22, 23]. Parallel designs include a ten-week pause in the study after half of all clusters had been vaccinated to allow for statistical analyses to be completed. We assume 80% vaccine coverage of susceptible individuals in targeted clusters, and that vaccine immediately removes individuals from a susceptible state 95% of the time. We continue the simulation until each epidemic has died out. We repeat the network generation and epidemic simulation process until we have 1,000 complete iterations. Parameter values for network generation, epidemic model, and infection and vaccination models are provided in Supplementary Table 2.

**Statistical analyses**

For each simulation we compute three metrics to quantify the epidemic outcomes: (1) time from epidemic start ($T_0$) until $R_e$ first falls below one; (2) time from $T_0$ until the last infectious individual recovers; and (3) cumulative incidence for the entire epidemic. These metrics are intended to evaluate the: (i) speed of control; (ii) speed of elimination; and (iii) overall burden of the epidemic. For each metric, we calculate median and interquartile range across all 1,000 simulations.

We compute statistical power to detect effectiveness of a vaccine using permutation tests based on pairwise comparisons of incidence across treatment and control clusters for the six study designs that involved randomization (Table 1). The test statistic for a null hypothesis of equal incidence rates in treatment and control clusters for week $T$ after the treated cluster was vaccinated, is the sum of the differences in incidence rate between clusters in each simulation, in week $T$; null hypotheses of equal incidence cumulative from week $T$ to week $T + w$ are similar. For Parallel designs, the test statistic is the sum of $K/2$ differences in incidence rates; for Stepped Wedge

Page 9

designs, it involves $(K - h - 1)$ differences. Each test involves 2000 permutations, and p-values are the proportion of test statistics greater or equal to the observed test statistic in absolute value. We calculate the probability of rejecting the null hypothesis at week $T - 2$ (i.e. two weeks prior to vaccination) and at each subsequent week up to $T + 10$. Our designs differ from usual CRTs in that pairs are formed based on their connectivity, not matched on potential cluster characteristics that are predictive of the outcome. This connectivity-based matching means that pairs are not strictly exchangeable, and thus our pair-wise randomization may not fully protect us from confounding when estimating treatment effects, if factors are associated both with connectedness and infection likelihood within clusters. The validity of the permutation tests is, however, guaranteed by randomization.

To assess sensitivity of results to key vaccine, trial and population characteristics, we conduct sensitivity analyses. First, we run a model for a vaccine with no effectiveness (for type-I error control). Second, we modeled vaccines that are: (i) perfect (100% reach and 100% protective); (ii) poor (70% reach and 70% effective); and (iii) able to protect those in the Exposed as well as the Susceptible state, moving them directly to the Removed state. Third, we began the vaccination program at 56 and 70 days post-initial infection. Fourth, we varied the heterogeneity in connectivity between clusters by decreasing and increasing the standard deviation of the between-cluster ties term from 0.5 to 0.25 and 0.75 contacts respectively. Finally, we considered clusters with skewed within-cluster ties by drawing each respondent's degree from a lognormal distribution with $\sigma = 1$. For each sensitivity analysis we summarized results in terms of the key metrics for epidemic outcomes.

**RESULTS**

In the absence of an intervention, the spreading process infects a median of 80.0% of the population (interquartile range [IQR] across 1,000 runs: 79.0-81.0%), peaks at a mean of 107.4 incident cases per 1000 susceptible individuals per week, and dies out after a median of 290 days (IQR 270-316). Mean $R_e$ declines slowly from a peak of 2.08 on day 36, falling below 1 after a median of 134 days (IQR: 120-147). Temporal plots of mean values for all state variables are shown in **Supplementary Figure 2**.

As intended, all study designs lead to substantial reductions in proportion of individuals ever infected, time to end of outbreak and time to $R_e$<1, relative to no vaccination (Table 2). All connectivity-informed designs lead to lower peak infectiousness than with traditional study designs, with little discernable difference among them (Figure 2B). Both of the Parallel designs suffer from the pause for analysis prior to vaccinating the second half of clusters, leading to lower levels of overall vaccine coverage (an average of 44.4% for Standard and 47.4% for Ranked Parallel vs. 52.3-56.1% for the Stepped Wedge designs) and higher overall mortality (an average of 23.2-25.5% for Parallel vs. 16.8-19.2% for Stepped Wedge designs) (Figure 2C and D). The effectiveness of Parallel designs prior to the pause is however, identical to their Stepped Wedge analogues, since the designs are identical prior to this point. Connectivity-informed Stepped Wedge designs reduce the proportion of individuals ever infectious relative to the Standard Stepped Wedge by approximately 20%, but do not have meaningful impact on time to last infection or to infection control (Table 2). The Ranked Parallel design similarly reduce the proportion ever infectious, compared to the Standard Parallel design, but is approximately one week slower to control the epidemic.

Cluster-level mean incidence rates decline as each cluster is vaccinated (Figure 3). In the Static Rank approach, the clusters that were treated last – and thus had lowest baseline connectivity – had

Page 11

notably lower incidence rates throughout the follow-up period; this was not the case for the other designs where the last-treated clusters were determined either at random (Standard approach) or based on Adaptive Ranks.

The Parallel designs show lower power to detect differences in incidence in the first week after vaccination; however ability to detect a difference increases monotonically over subsequent periods and at the tenth week post-treatment, these designs are the most powerful of all (Figure 4A). Power rises most rapidly for the Standard Stepped Wedge design; however after four weeks it begins to fall as control clusters begin to become treated. The effect of treatment of control clusters is even stronger for Fuzzy Order designs, where power peaks at 66.2% (Static Rank) and 64.1% (Adaptive Rank) at three weeks post-treatment, and fall back to non-significantly different from 5% by week 10. The Holdback-1 design falls between the Standard and Fuzzy Order designs, peaking at 87.8% power three weeks post-vaccination and falling off more slowly. Using cumulative post-treatment incidence, all connectivity-informed designs show at least 60% power based on the first 21 days of post-treatment person-time (Figure 4B).

Sensitivity analyses suggest that time to epidemic end is affected strongly by the effectiveness of the vaccine (such that a vaccine protecting only 50% of cluster members leads to outcomes almost equal to no vaccine at all), but only weakly by the timing of vaccine initiation (**Supplementary Table 3**). Cumulative incidence is substantially reduced under all vaccine effectiveness scenarios. While vaccine quality remains important, later trial initiation leads to substantially increased cumulative incidence. The probability of detecting a significant difference between cluster incidence in the first week post-treatment varies greatly depending on the effectiveness of the vaccine for all designs. For the Parallel designs, week of trial start also affected power, although this was not the case for Stepped Wedge designs. Varying the nature of connectivity between and within clusters did not substantially affect any outcome within the bounds we considered.

**DISCUSSION**

This paper proposes a class of connectivity-informed designs for cluster randomized trials that provide more rapid epidemic control in return for reduced power to detect vaccine effects. This potential improvement in trial design for epidemic settings is made possible by leveraging knowledge of the contact network between clusters. Based solely on a relative ordering of how connected members of a cluster are to the outside world (Static Rank) or absolute levels of connectivity (Adaptive Rank), our designs aim to minimize the potential number of cross-cluster infections that can arise, and thus reduce the rate at which incidence grows and the overall epidemic size. In illustrative simulations, we show that such designs reduce the total number of infections in the population by up to 20%, compared to traditional staggered CRTs. This benefit appears to be spread across clusters treated both early and late in the trial (Figure 3). In any trial of an intervention that is expected to reduce disease burden, we desire to: (i) generate a valid test of effect; (ii) generate a public health benefit; and (iii) get randomized effect estimates. Our class of designs achieves the first two of these goals, at the cost of the third.

Within the class of designs we propose, the relative level and timing of each study's ability to detect a significant difference between study arms varies. The upward sloping power curves in Figure 4 reflect the direct effect of vaccination, so that a case prevented in one time period has follow-on benefits in subsequent periods since that case is now not infecting others [8]. All of the connectivity-informed designs perform worse than their connectivity-naïve counterparts, even in the period immediately following vaccination, at least in part due to the lower overall incidence seen in these scenarios. Low power a few weeks post-vaccination is expected for the Fuzzy Order designs, as the control cluster at one time point has a 50% chance of being vaccinated at the following time point, reducing the likely difference in incidence rates between arms. This effect can be seen in the later weeks following vaccination for the Fuzzy Order designs with no holdback period. Adding a holdback period (i.e. a period during which each control cluster cannot be vaccinated), however, greatly improved study power in our simulation – regaining more than two-thirds of the power lost by shifting from a Standard to Fuzzy Order design, with little impact on infection levels. A brief holdback period therefore appears wise in settings where insufficient sample sizes can be obtained to ensure high power.

The choice of design within our connectivity-informed class will depend on the study goals. A Strict Order design should be optimal for epidemic control, since it leads to the fastest immunization of potentially-infectious ties between clusters; but it does not provide randomized comparisons. A Fuzzy Order design does permit randomization; however it can only ensure that each control cluster remains untreated for a single period which, depending on how fast the infection spreads, may not be sufficient for substantial differences between treatment and control arms to emerge. Adding $h$ holdback periods to the Fuzzy Order design offers the assurance that control clusters remain untreated for longer, which should allow larger differentials in incidence to arise. The trade-off is that some highly connected clusters may not be treated early, and $h$ statistical comparisons will be lost.

A comparison of our designs to the Ring Vaccination approach is illustrative. Both methods leverage network structure – in the case of Ring Vaccination based on networks of individuals, in our case based on networks of clusters. When there are few cases, and for a disease that is highly symptomatic while infectious, Ring Vaccination should ensure very rapid control. When the number of cases is large, individual-level attention is likely to be either too resource-intensive or pointless: if most people have a contact or contact-of-contact who is infected, mass treatment may well be more efficient than contact-tracing [24]. In contrast, our designs are likely to be most beneficial in a widespread epidemic where infectious individuals are hard to identify, and case rates are rising rapidly.

As with any study, ours has limitations. First, our methods require information on connectivity, preferably on the ties through which epidemics are passed (e.g. sexual partners, close contact). This information is not always easy to gather, however more-easily measured quantities (e.g. distance, transport links or cellphone call rates) might be used as proxies, and one of our two approaches requires only a rank ordering of how connected clusters are to the outside world. Second, as our designs are complex, analytic results are not easily available. The specific forms of our simulations are only illustrative, and the impact of our designs on epidemic outcomes and power measurements will depend on both network characteristics and infection dynamics. Context-specific simulation studies are therefore likely to be needed to determine sample sizes required using these methods, and to determine their usefulness relative to Standard approaches. Third, our designs do not provide a straightforward way to estimate vaccine effectiveness, due in part to the complication that assignment of a cluster to treatment or control depends on prior randomizations. As a result, there is a possibility of confounding by any factors that predict both cluster-level connectivity and infection risk. Our models may therefore be most useful in settings where maximizing public health impact is more important than establishing precise effectiveness.

This paper can be considered illustrative, and thus could be extended in several ways. First, we have presented a non-exhaustive set of exemplar study designs from within a broad connectivity-informed class. Connectivity information could be used in various other ways. For example, connectivity could be measured relative to an outbreak source, rather than equally to all other clusters. Alternatively, one might wish to account for the internal connectivity structure of clusters – either in addition or instead of between-cluster connections. Second, we have presented a vaccine candidate for a specific infection; our approach could be extended to other infections, the spread of ideas or behaviors or non-vaccine treatments. The only requirements for our methods are the existence of a spreading process and an intervention that affects this process. Third, we have presented a simple case in which the study population is homogeneous except for each person's number of contacts. The approach can be extended to adjust for individual covariates of individuals, or characteristics (e.g. strength) of the ties between individuals.

We show here that integrating information about cluster connectivity can strengthen CRT designs, especially in acute outbreak settings. While the exact benefits will differ depending on the infection and the social context of the outbreak, we suggest that connectivity-informed designs may play an important role in the implementation process of novel interventions, even if they are unable to answer all questions required to confirm the usefulness of a new vaccine or treatment.

**Acknowledgements**: We thank participants at the 2015 Ebola Modeling Workshop at Georgia Tech, at Healthmap and in the Onnela lab group for comments on presentations of these ideas. We thank Laura Balzer for her several helpful comments on a draft of this paper.

**Disclaimer**: The funders had no role in study design, data collection and analysis, decision to publish, or preparation of the manuscript.

**Authors' contributions**: All authors contributed to the study conception and design, data interpretation and final revisions to the text. GH conducted the network simulations and statistical analyses with contributions from RW and JPO. GH summarized the results in tables and graphs and wrote the first draft of the paper.

**Declaration of Conflicting Interests**: The Authors declare that there is no conflict of interest.

**SUPPLEMENTARY MATERIAL**

**Title**: Leveraging contact network structure in the design of cluster randomized trials

**FIGURES AND TABLES**

**Figure 1: Between-cluster connectivity calculated at different times in a cluster randomized trial**

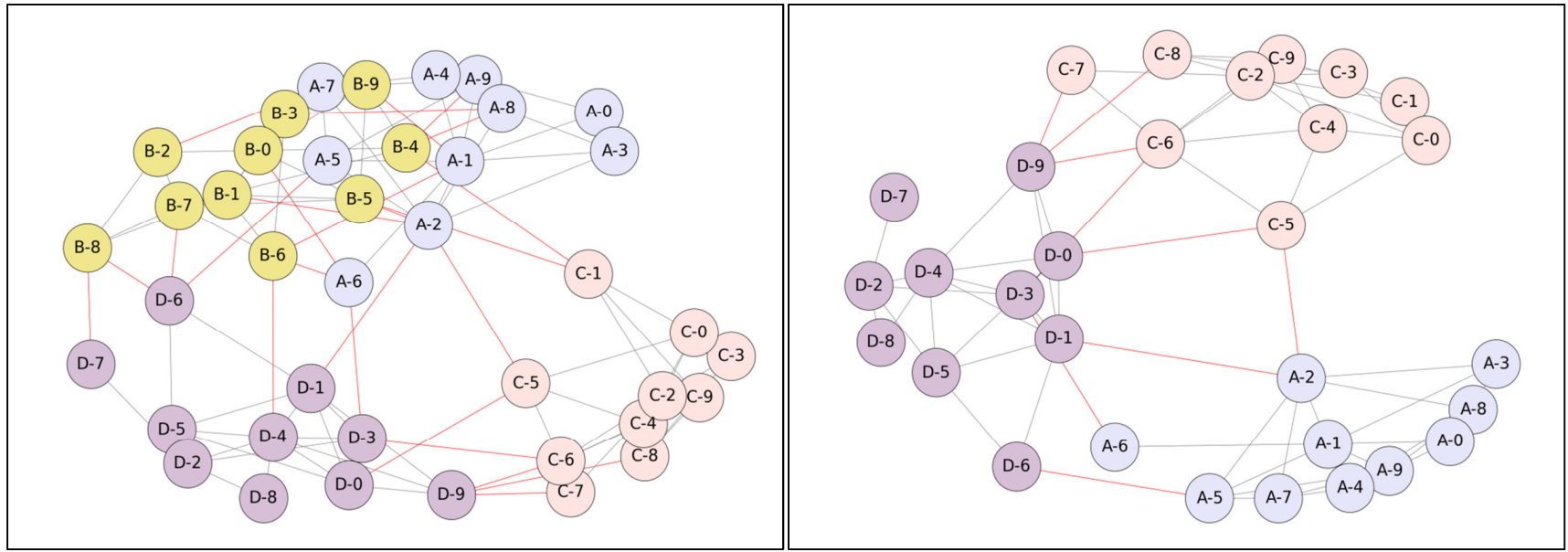

A                                                                                                                 B

Each figure shows four clusters, each containing 10 individuals. Within-cluster ties are shown in grey (paler) and between-cluster ties are shown in red (darker). In panel A, cluster A has 14 ties to other clusters; cluster B 16; cluster C 8 and cluster D 12; the ordering of cross-cluster ties is thus (B, A, D, C). The Static Rank Strict Order design will follow the order B, A, D, C. The Static Rank Fuzzy Order design will first treat either A or B, with the other acting as control; at step 2, the untreated cluster from A or B will be randomized against D; and at step 3 the remaining untreated cluster will be randomized against cluster C. The Static Rank Fuzzy Order Holdback-1 would first treat either A or B, with the other acting as control and then being barred from randomization at step 2; at step 2 C and D would be randomized; at step 3 only the untreated cluster from the A and B randomization would be available and thus treated; at step 4 the final cluster would be treated.

In panel B, after cluster 'B' has been treated and removed from consideration, cluster 'A' has moved from the second most-connected to the least-connected cluster; the ordering has now changed to (D, C, A).

**Figure 2: Mean state values for each day since the start of the epidemic across 1000 simulations, selected states**

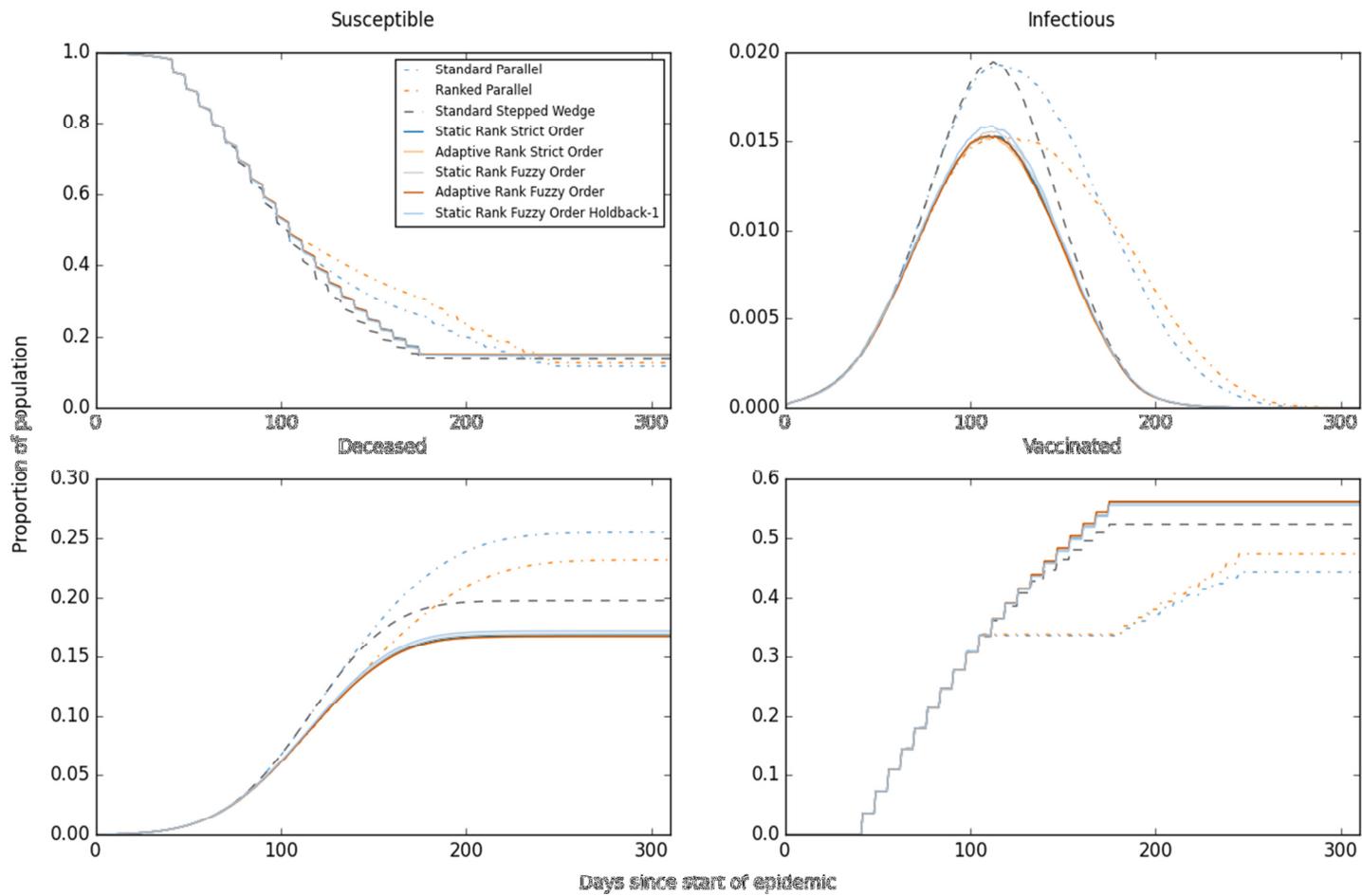

**Figure 3: Mean weekly incidence (per 1000 susceptible individuals) within clusters**

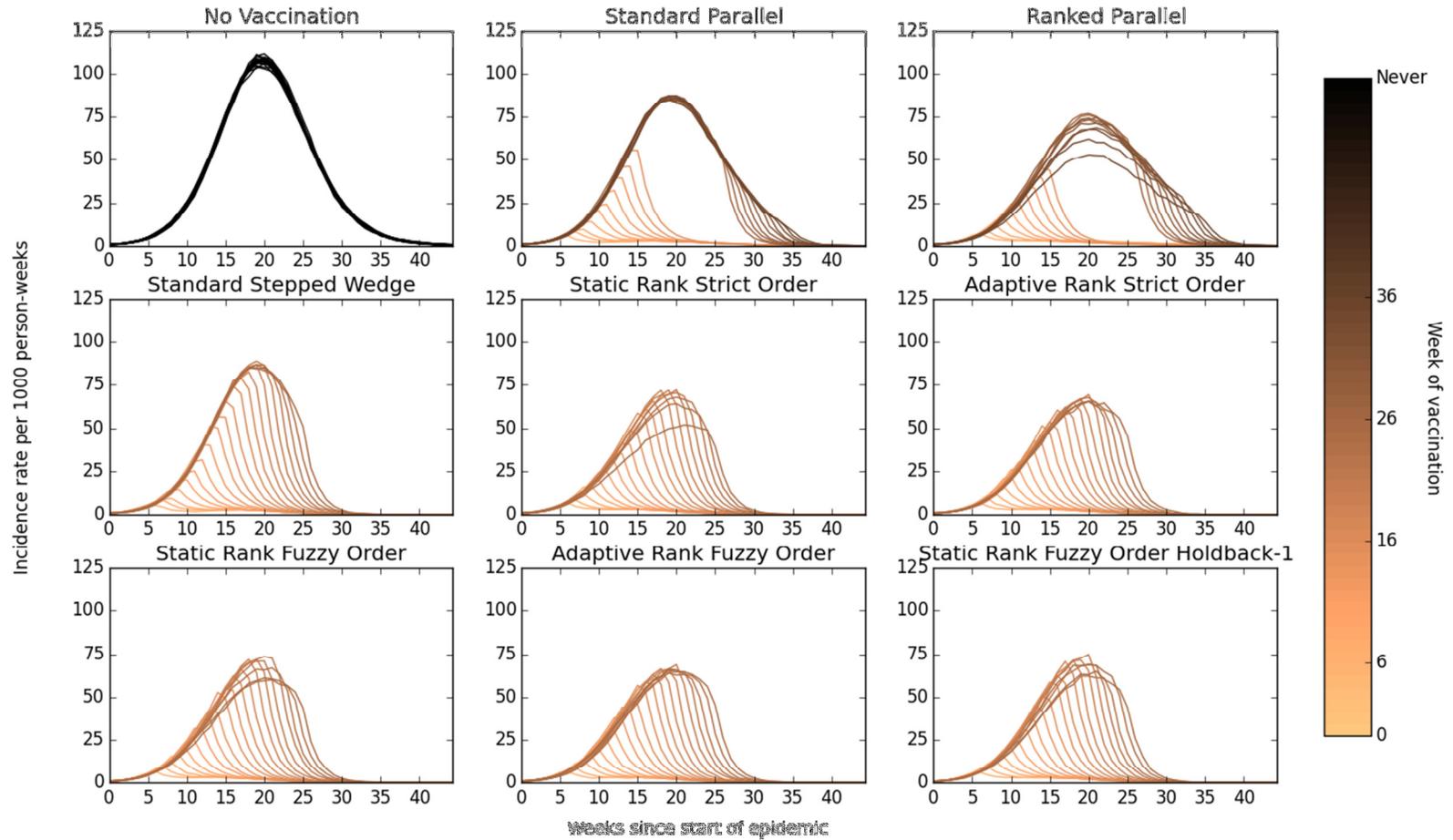

Darker lines represent earlier randomization times, thus the darkest line represents the mean incidence rate in clusters that were treated at the first possible time point (week 6) in each simulation realization.

# Figure 4: Power of study designs to detect a significant effect of the vaccine

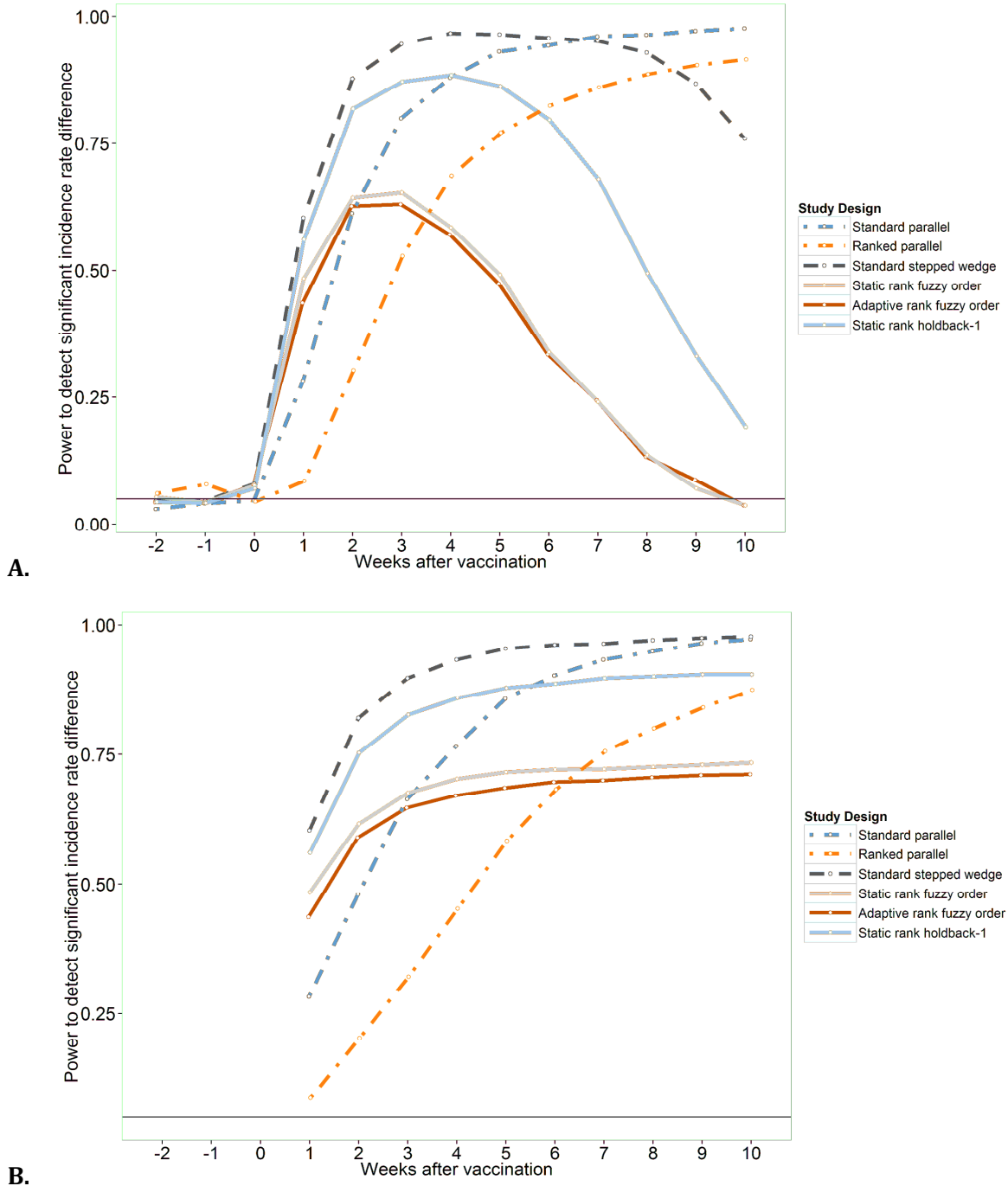

A. Marginal incidence in each week; B. Cumulative incidence since treatment.

Figures show the proportion of 1000 simulations able to detect a significant difference at α=0.05 between the treatment and control clusters across $K/2$ (Standard Parallel and Ranked Parallel), $K-1$ (Stepped Wedge, Static Ranked Fuzzy, Adaptive Ranked Fuzzy) or $K-2$ (Static Ranked Holdback) permutation tests based on pairwise comparisons of incidence in the week specified relative to week of vaccination of the treatment cluster in each pair. Permutation test based on 2000 permutations. Data on which these figures are based are shown in Supplementary Table 1.

Page 24

**Table 1: A typology of study designs used in this paper**

|  | Treatment assignment [a] | Time of ranking | Treatment timing [b] | Number of comparisons for power calculation [c] |
|---|---|---|---|---|
| No Vaccination | - | - | - | - |
| Standard Parallel | Unranked PR | Never | Pause halfway | $K/2$ |
| Ranked Parallel | Ranked PR | At baseline | Pause halfway | $K/2$ |
| Standard Stepped Wedge | Randomized order | Never | Continuous | $K-1$ |
| Static Rank Strict Order | Strictly by rank | At baseline | Continuous | - |
| Static Rank Fuzzy Order | Ranked PR | At baseline | Continuous | $K-1$ |
| Static Rank Fuzzy Order Holdback-1 [d] | Ranked PR | At baseline | Continuous | $K-2$ |
| Adaptive Rank Strict Order | Strictly by rank | At each step | Continuous | - |
| Adaptive Rank Fuzzy Order | Ranked PR | At each step | Continuous | $K-1$ |

[a] PR: Pair randomization.
[b] "Pause halfway" designs are those where one cluster in each pair is treated sequentially, and then after an evaluation period, if the vaccine proves protective, continuing to treat the other clusters sequentially. "Continuous" designs treat one new cluster at each vaccination time point until all clusters are treated.
[c] Strict Order designs do not evaluate vaccine effectiveness.
[d] In this design the untreated cluster in each randomized pair is not eligible for randomization at the next vaccination time step.

**Table 2: Population-level spreading process outcomes** [a]

|  | Proportion ever infectious (%) | Time to last infectious individual (days) | Time to $R_e < 1$ (days) [b] |
|---|---|---|---|
| No randomization | | | |
|   No Vaccination | 80.1 [79.0 - 81.0] | 138 [127 - 152] | 290 [270 - 316] |
|   Static Rank Strict Order | 30.1 [20.5 - 37.7] | 119 [109 - 130] | 228 [216 - 243] |
|   Adaptive Rank Strict Order | 29.6 [19.8 - 38.1] | 119 [109 - 130] | 228 [217 - 241] |
| Parallel designs | | | |
|   Standard Parallel | 44.9 [41.3 - 48.5] | 136 [118 - 156] | 265 [251 - 281] |
|   Ranked Parallel | 42.2 [34.7 - 46.9] | 142 [115 - 164] | 278 [262 - 293] |
| Stepped Wedge designs | | | |
|   Standard Stepped Wedge | 35.4 [27.6 - 41.5] | 121 [111 - 132] | 228 [218 - 241] |
|   Static Rank Fuzzy Order | 30.0 [21.3 - 38.1] | 121 [110 - 130] | 229 [217 - 243] |
|   Static Rank Fuzzy Order Holdback-1 | 30.4 [21.7 - 38.0] | 119 [109 - 130] | 228 [218 - 242] |
|   Adaptive Rank Fuzzy Order | 29.4 [20.3 - 38.0] | 118 [108 - 130] | 228 [217 - 243] |

[a] All outcome figures are medians and interquartile ranges of result from 1000 simulation realizations.
[b] $R_e$: effective reproductive number.

**Supplementary Material**

**Title:** Leveraging contact network structure in the design of cluster randomized trials

**Supplementary Table 1: Probability of rejecting null hypothesis of no effect of intervention comparing incidence in treatment cluster to control cluster**

**Marginal: incidence in the week indicated**

|  | T-2 | T-1 | T | T+1 | T+2 | T+3 | T+4 | T+5 | T+6 | T+7 | T+8 | T+9 | T+10 |
|---|---|---|---|---|---|---|---|---|---|---|---|---|---|
| Parallel designs | | | | | | | | | | | | | |
|   Standard Parallel | 3.0 | 4.1 | 4.7 | 28.2 | 61.2 | 79.9 | 87.9 | 93.1 | 94.3 | 96.1 | 96.4 | 97.2 | 97.7 |
|   Ranked Parallel | 6.1 | 7.9 | 4.5 | 8.6 | 30.3 | 52.8 | 68.7 | 77.1 | 82.4 | 86.0 | 88.6 | 90.4 | 91.5 |
| Stepped Wedge designs | | | | | | | | | | | | | |
|   Standard Stepped Wedge | 4.6 | 4.6 | 8.0 | 60.3 | 87.7 | 94.6 | 96.7 | 96.5 | 95.8 | 95.2 | 92.9 | 86.7 | 75.9 |
|   Static Rank Fuzzy Order | 5.4 | 4.3 | 7.7 | 48.4 | 64.3 | 65.3 | 58.4 | 49.1 | 33.9 | 24.1 | 13.6 | 7.1 | 3.7 |
|   Static Rank Fuzzy Order Holdback-1 | 4.4 | 4.2 | 7.1 | 56.0 | 81.8 | 87.1 | 88.4 | 86.2 | 79.6 | 68.0 | 49.4 | 33.1 | 19.2 |
|   Adaptive Rank Fuzzy Order | 4.3 | 4.2 | 7.8 | 43.6 | 62.6 | 63.0 | 57.0 | 47.4 | 33.5 | 24.4 | 13.3 | 8.8 | 3.7 |

**Cumulative: incidence in all weeks from treatment up to the week indicated**

|  | T-2 | T-1 | T | T+1 | T+2 | T+3 | T+4 | T+5 | T+6 | T+7 | T+8 | T+9 | T+10 |
|---|---|---|---|---|---|---|---|---|---|---|---|---|---|
| Parallel designs | | | | | | | | | | | | | |
|   Standard Parallel | | | | 28.2 | 48.0 | 66.4 | 76.6 | 85.7 | 90.2 | 93.3 | 94.9 | 96.5 | 97.3 |
|   Ranked Parallel | | | | 8.6 | 20.3 | 32.1 | 45.2 | 58.3 | 68.3 | 75.6 | 80.0 | 84.0 | 87.4 |
| Stepped Wedge designs | | | | | | | | | | | | | |
|   Standard Stepped Wedge | | | | 60.3 | 82.0 | 89.7 | 93.3 | 95.5 | 96.2 | 96.4 | 97.1 | 97.5 | 97.8 |
|   Static Rank Fuzzy Order | | | | 48.4 | 61.6 | 67.6 | 70.3 | 71.6 | 72.1 | 72.2 | 72.6 | 73.0 | 73.4 |
|   Static Rank Fuzzy Order Holdback-1 | | | | 56.0 | 75.3 | 82.6 | 85.8 | 87.8 | 88.6 | 89.7 | 90.0 | 90.4 | 90.4 |
|   Adaptive Rank Fuzzy Order | | | | 43.6 | 58.9 | 64.7 | 67.1 | 68.6 | 69.7 | 70.0 | 70.6 | 71.0 | 71.2 |

All dates are relative to week $T$, the week of treatment in the treatment cluster. Figures are power to reject a null hypothesis of no difference between the treatment and control cluster. For Parallel designs probability is calculated across 10 paired comparisons; for Stepped Wedge designs probability is calculated across 19 paired comparisons except for Static Rank Order Holdback-1 with 18 paired comparisons. In all cases, the probability is calculated using a permutation test based on pairwise comparisons of incidence in 1000 simulations, with 2000 permutations per test.

**Supplementary Table 2: Parameter values for primary simulations**

| | |
|---|---|
| Cluster characteristics [†] | |
|   Number of clusters | 20 |
|   Number of individuals per clusters | 200 |
|   Mean (standard deviation) within-cluster degree | 4.5 (0) & 5 (0) |
|   Mean (standard deviation) between-cluster degree | 1 (0.5) & 0.5 (0.5) |
| | |
| Infection characteristics | |
|   Number of initial infections | 4 |
|   Per-day risk of transmission from contact: | |
|   Infectious at home ($\beta_I$) | 7.5% |
|   Hospitalized ($\beta_H$) | 3.75% |
|   Deceased, not buried ($\beta_F$) | 10% |
|   Mean duration in state (days): | |
|   Incubation period ($\gamma_E^{-1}$) | 9 |
|   From symptom onset to hospitalization ($\gamma_H^{-1}$) | 5 |
|   From symptom onset to recovery ($\gamma_I^{-1}$) | 10 |
|   From symptom onset to death ($\gamma_D^{-1}$) | 10 |
|   From hospitalization to recovery ($\gamma_{IH}^{-1}$) | 5 |
|   From hospitalization to death ($\gamma_{DH}^{-1}$) | 5 |
|   From death to burial ($\gamma_F^{-1}$) | 2 |
|   Proportion of cases hospitalized ($\theta$) | 0.5 |
|   $\theta_I$ | 0.33 |
|   Case-fatality rate: | |
|   Without hospitalization ($\delta_I$) | 0.75 |
|   With hospitalization ($\delta_H$) | 0.65 |
| | |
| Vaccination characteristics | |
|   Time between vaccination rounds | 7 days |
|   Clusters vaccinated per round | 1 |
|   Pause period for evaluation, Parallel trials | 70 days |
|   Proportion of cluster residents successfully vaccinated | 80% |
|   Probability of vaccination removing individual | 95% |

[†] The 20 study clusters were divided into two groups; both had the same mean number of contacts per person, but individuals in clusters in one group had (on average) twice as many between-cluster contacts as the other.

**Supplementary Table 3: Key epidemic and study outcomes for sensitivity analyses**

|  | Standard Parallel | | Ranked Parallel | | Standard Stepped Wedge | | Static Rank Strict Order | | Static Rank Fuzzy Order | | Static Rank Holdback-1 | |
|---|---|---|---|---|---|---|---|---|---|---|---|---|
| Time to end of epidemic | | | | | | | | | | | | |
| Baseline | 265 | [251 - 281] | 228 | [217 - 241] | 228 | [218 - 241] | 228 | [216 - 243] | 229 | [217 - 243] | 228 | [218 - 242] |
| Negative control | 292 | [272 - 319] | 291 | [269 - 317] | 291 | [272 - 317] | 293 | [274 - 320] | 291 | [273 - 315] | 293 | [273 - 320] |
| Poor vaccine | 292 | [269 - 322] | 290 | [264 - 331] | 281 | [255 - 312] | 290 | [263 - 323] | 293 | [264 - 330] | 289 | [262 - 327] |
| Perfect vaccine | 257 | [245 - 270] | 209 | [201 - 217] | 209 | [203 - 219] | 207 | [199 - 216] | 208 | [200 - 216] | 209 | [201 - 218] |
| Perfect vaccine incl. Exposed | 257 | [243 - 271] | 219 | [207 - 234] | 219 | [208 - 233] | 219 | [206 - 232] | 218 | [205 - 234] | 221 | [208 - 234] |
| Week 8 vaccine | 267 | [252 - 282] | 237 | [227 - 251] | 234 | [223 - 248] | 238 | [227 - 250] | 238 | [225 - 251] | 237 | [226 - 249] |
| Week 10 vaccine | 270 | [254 - 286] | 244 | [232 - 256] | 239 | [228 - 251] | 245 | [234 - 258] | 244 | [233 - 259] | 244 | [233 - 257] |
| Low between-heterogeneity | 267 | [252 - 284] | 231 | [221 - 244] | 229 | [217 - 242] | 230 | [218 - 243] | 230 | [218 - 245] | 231 | [219 - 244] |
| High between-heterogeneity | 262 | [248 - 279] | 228 | [215 - 241] | 227 | [215 - 240] | 227 | [215 - 239] | 226 | [215 - 241] | 227 | [215 - 241] |
| Lognormal within-cluster ties | 266 | [250 - 283] | 228 | [218 - 242] | 228 | [216 - 242] | 229 | [219 - 243] | 228 | [218 - 242] | 229 | [218 - 242] |
| Cumulative incidence | | | | | | | | | | | | |
| Baseline | 44.9 | [41.3 - 48.5] | 29.6 | [19.8 - 38.1] | 35.4 | [27.6 - 41.5] | 30.1 | [20.5 - 37.7] | 30.0 | [21.3 - 38.1] | 30.4 | [21.7 - 38.0] |
| Negative control | 80.0 | [78.8 - 81.1] | 79.9 | [78.8 - 80.9] | 80.0 | [78.9 - 81.0] | 79.8 | [78.8 - 81.1] | 79.9 | [78.8 - 81.0] | 80.0 | [78.8 - 81.1] |
| Poor vaccine | 51.9 | [48.6 - 55.1] | 38.9 | [30.7 - 46.4] | 44.0 | [37.3 - 49.8] | 40.2 | [31.0 - 47.4] | 39.3 | [30.9 - 47.4] | 40.1 | [31.2 - 47.7] |
| Perfect vaccine | 42.2 | [38.2 - 45.4] | 25.2 | [17.2 - 32.5] | 30.5 | [22.8 - 37.4] | 25.2 | [17.0 - 33.2] | 25.1 | [17.8 - 33.7] | 25.8 | [17.5 - 33.5] |
| Perfect vaccine incl. Exposed | 43.0 | [39.1 - 46.3] | 23.0 | [15.6 - 31.1] | 29.9 | [22.0 - 36.9] | 24.0 | [15.7 - 31.4] | 24.6 | [15.8 - 32.3] | 24.7 | [17.3 - 32.7] |
| Week 8 vaccine | 49.6 | [45.5 - 53.5] | 39.6 | [30.2 - 47.0] | 43.0 | [34.9 - 49.1] | 40.1 | [30.1 - 47.3] | 40.4 | [30.2 - 48.2] | 40.0 | [29.2 - 48.5] |
| Week 10 vaccine | 54.5 | [50.0 - 59.5] | 50.4 | [39.6 - 57.4] | 51.5 | [44.3 - 57.8] | 50.0 | [40.6 - 57.9] | 49.6 | [39.6 - 57.1] | 50.0 | [40.9 - 57.6] |
| Low between-heterogeneity | 44.5 | [41.0 - 48.0] | 30.7 | [22.7 - 37.6] | 34.8 | [27.1 - 40.6] | 31.1 | [22.2 - 37.9] | 30.7 | [22.0 - 37.7] | 31.5 | [23.1 - 38.6] |
| High between-heterogeneity | 45.4 | [41.2 - 49.2] | 29.4 | [19.8 - 38.0] | 36.5 | [27.8 - 42.8] | 28.4 | [19.4 - 37.9] | 29.0 | [18.9 - 38.2] | 29.2 | [20.2 - 37.8] |
| Lognormal within-cluster ties | 45.0 | [40.7 - 48.4] | 28.8 | [20.3 - 37.5] | 35.4 | [27.4 - 41.5] | 28.8 | [19.7 - 36.9] | 29.2 | [20.9 - 38.1] | 29.4 | [21.0 - 38.0] |
| Power after one week | | | | | | | | | | | | |
| Baseline | 27.3 | [24.5 - 30.1] | 7.8 | [6.1 - 9.5] | 60.6 | [57.6 - 63.6] | | | 46.8 | [43.7 - 49.9] | 54.1 | [51.0 - 57.2] |
| Negative control | 4.1 | [2.9 - 5.3] | 10.2 | [8.3 - 12.1] | 4.9 | [3.6 - 6.2] | | | 4.6 | [3.3 - 5.9] | 4.5 | [3.2 - 5.8] |
| Poor vaccine | 13.5 | [44.6 - 50.8] | 4.5 | [3.2 - 5.8] | 32.3 | [29.4 - 35.2] | | | 24.8 | [22.1 - 27.5] | 30.1 | [27.3 - 32.9] |
| Perfect vaccine | 47.7 | [11.4 - 15.6] | 16.1 | [13.8 - 18.4] | 80.7 | [78.3 - 83.1] | | | 62.4 | [59.4 - 65.4] | 70.4 | [67.6 - 73.2] |
| Perfect vaccine incl. Exposed | 55.1 | [52.0 - 58.2] | 25.9 | [23.2 - 28.6] | 88.8 | [86.8 - 90.8] | | | 71.1 | [68.3 - 73.9] | 86.8 | [84.7 - 88.9] |
| Week 8 vaccine | 40.4 | [37.4 - 43.4] | 13.4 | [11.3 - 15.5] | 63.6 | [60.6 - 66.6] | | | 60.3 | [57.3 - 63.3] | 68.4 | [65.5 - 71.3] |
| Week 10 vaccine | 50.7 | [47.6 - 53.8] | 26.8 | [24.1 - 29.5] | 63.4 | [60.4 - 66.4] | | | 60.5 | [57.5 - 63.5] | 70.0 | [67.2 - 72.8] |
| Low between-heterogeneity | 27.6 | [24.8 - 30.4] | 10.0 | [8.1 - 11.9] | 60.4 | [57.4 - 63.4] | | | 47.8 | [44.7 - 50.9] | 56.8 | [53.7 - 59.9] |
| High between-heterogeneity | 31.3 | [28.4 - 34.2] | 5.8 | [4.4 - 7.2] | 58.5 | [55.4 - 61.6] | | | 46.7 | [43.6 - 49.8] | 51.7 | [48.6 - 54.8] |
| Lognormal within-cluster ties | 25.7 | [23.0 - 28.4] | 8.3 | [6.6 - 10.0] | 59.7 | [56.7 - 62.7] | | | 46.9 | [43.8 - 50.0] | 53.2 | [50.1 - 56.3] |

## Supplementary Figure 1: Schematic of state transition model

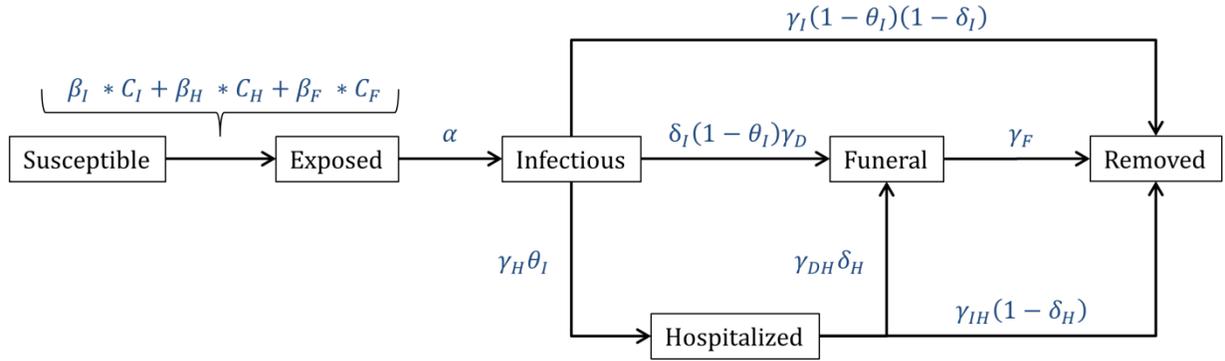

$\beta_I, \beta_H$ and $\beta_F$ are the transmission risk to a susceptible individual if one of the their contacts is, respectively, infectious at home ($C_I$), hospitalized ($C_H$) or deceased but not yet buried ($C_F$).
The inverse of each $\gamma$ term reflects the mean duration in a state: incubation period ($\gamma_E$); time from symptom onset to hospitalization ($\gamma_H$); time from symptom onset to recovery without hospitalization ($\gamma_I$); time from symptom onset to death without hospitalization ($\gamma_D$); time from hospitalization to recovery ($\gamma_{IH}$); time from hospitalization to death ($\gamma_{DH}$); and time to burial from death ($\gamma_F$).
$\theta_I$ is computed to fix the proportion of individuals hospitalized, allowing for competing risks of death or recovery, $\delta_I$ and $\delta_H$ are computed to fix the overall case-fatality ratio. Details, including equations for $\theta_I, \delta_I$ and $\delta_H$, can be found in Tables 2 and 3 of [21].

**Supplementary Figure 2: Mean state values for each day since the start of the epidemic across 1000 simulations**

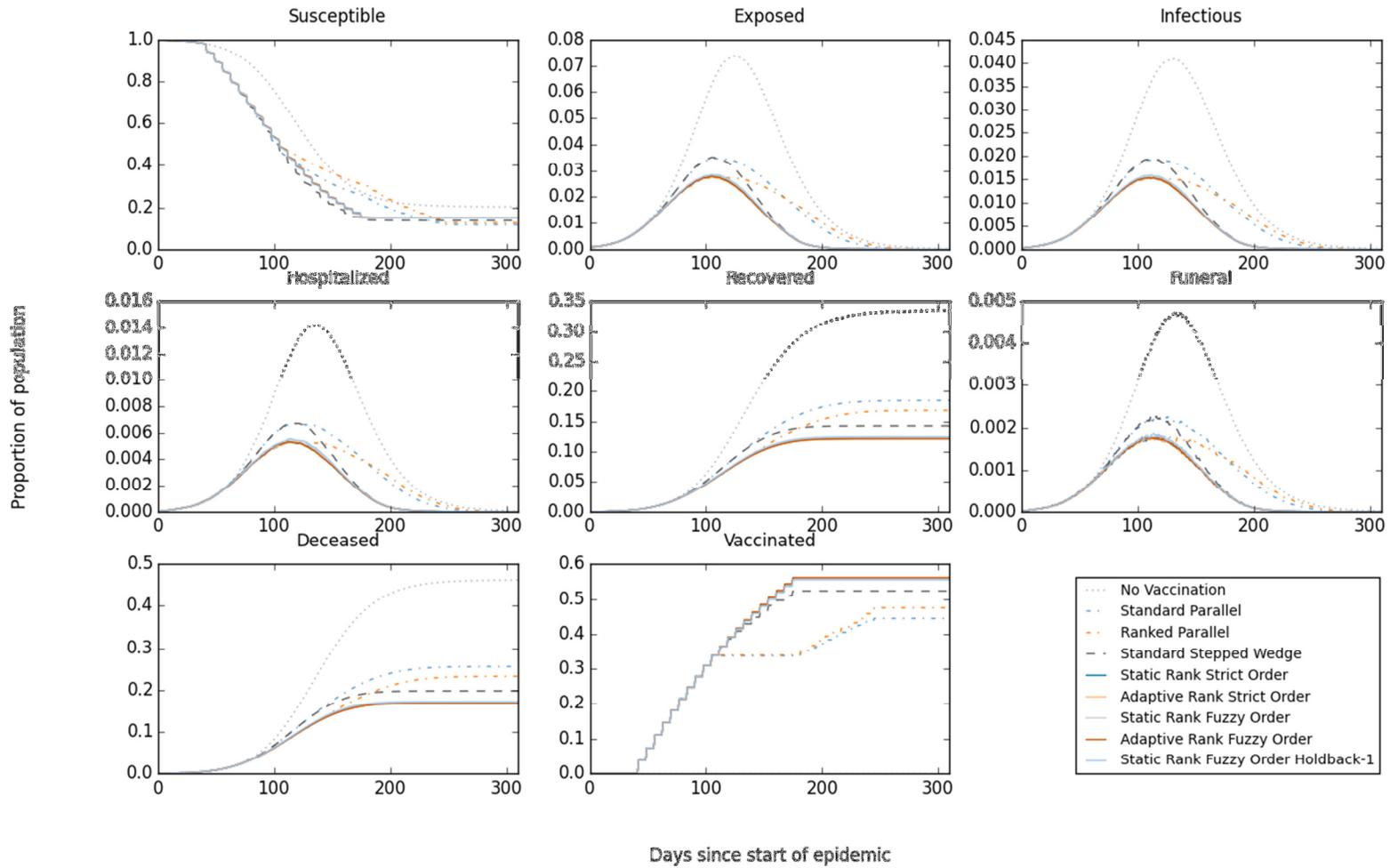

**Supplementary Figure 3: Mean daily effective reproductive rate for each vaccination study design across 1000 simulations**

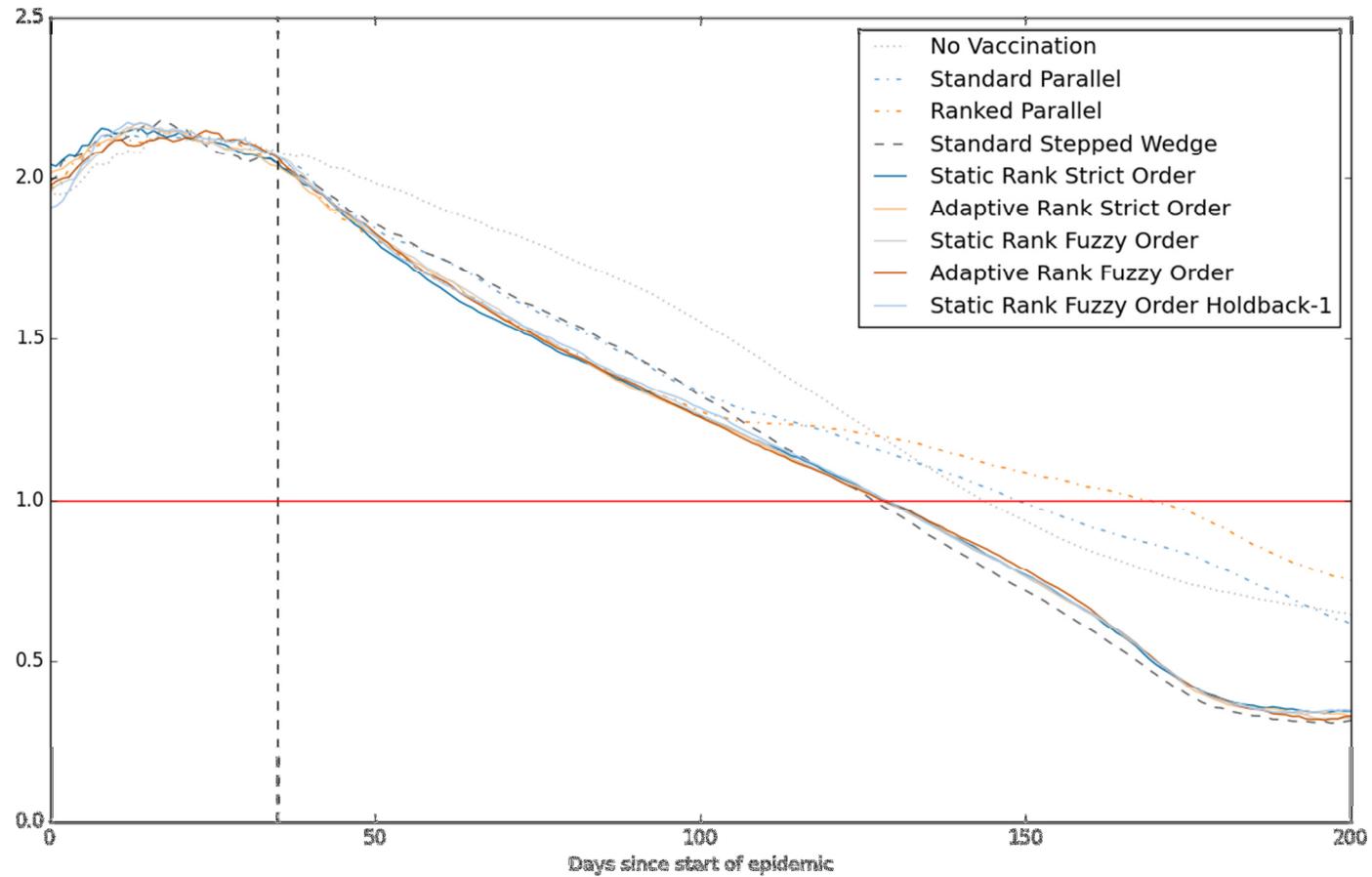

Vertical dashed line represents date of first cluster vaccination; horizontal solid line represents $R_e=1$.